\documentclass{aa}
\usepackage[dvips]{graphicx}
\def\simless{\mathbin{\lower 3pt\hbox
{$\rlap{\raise 5pt\hbox{$\char'074$}}\mathchar"7218$}}}   
\def\simmore{\mathbin{\lower 3pt\hbox
{$\rlap{\raise 5pt\hbox{$\char'076$}}\mathchar"7218$}}}   
\newcommand{\be}{\begin{equation}}
\newcommand{\ee}{\end{equation}}

\begin{document}
\title{Prompt emission spectra from the photosphere of a GRB}
\titlerunning{Prompt emission spectra from the photosphere of a GRB}
\author{Dimitrios Giannios}

\institute{Max Planck Institute for Astrophysics, Box 1317, D-85741 Garching, Germany}

\offprints{giannios@mpa-garching.mpg.de}
\date{Received / Accepted}

\abstract
{I explore the observational appearance of the photosphere of 
an ultrarelativistic flow with internal dissipation of energy (``dissipative'' GRB model). 
As a case study, I use the magnetic reconnection model (AC model)
that makes robust predictions on the energy dissipation rates at different 
radii in the flow. With analytical and numerical tools for the radiative
transfer problem, I show that the flow
develops a hot photosphere where inverse Compton scattering leads to highly
non-thermal spectrum. For a wide range of luminosities
and baryon loadings of the flow, this spectrum is very close to 
the observed prompt GRB emission. Its luminosity ranges from $\sim 3$
to 20\% of that of the total energy input.    
   
}

\maketitle

\section{Introduction} 
\label{intro}

Although much progress has been made the recent years in our understanding
 the central engine and of the afterglow emission of 
Gamma-ray bursts (GRBs), much less is known about the physical 
processes responsible for the prompt emission. Theoretical 
arguments related to the so called compactness problem
(e.g. Piran 1999) suggest that the emitting material is ultrarelativistic 
with bulk Lorentz factors $\Gamma\simmore 100$. The typical observed GRB 
spectrum has a characteristic peak (in $E\cdot f(E)$ representation) in the sub-MeV
range that is smoothly connected to low and high frequency power-laws, 
usually modeled with the Band spectrum (Band et al. 1993; Preece et al. 1998). 
Models for the prompt GRB phase have to account for both the acceleration 
of the flow and its observed spectral properties.       

For a flow to be accelerated to high bulk Lorentz factors, it must start
with high energy-to-rest-mass ratio. Depending on whether the energy
is in thermal or magnetic form, one has a fireball (Paczynski 1986; Goodmann 1986)
or a Poynting-flux dominated flow (Thompson 1994; M\'esz\'aros \& Rees 1997; Spruit et al.
2001; Drenkhahn \& Spruit 2002; Lyutikov \& Blandford 2003). In the fireball model, the flow
passes through an initial phase of rapid acceleration where $\Gamma\sim r$ until most of the 
internal energy has been used to accelerate the flow (unless the baryon 
loading is very low and radiation decouples from matter before the acceleration
phase is over). Further out, internal shocks (Rees \&  M\'esz\'aros 1994; Sari \& 
Piran 1997) can dissipate part of the kinetic energy and power the prompt emission.
In magnetic models for GRBs, the flow acceleration is more gradual (e.g. 
Drenkhahn \& Spruit 2002; Vlahakis \& K\"onigl 2003) and dissipation of magnetic
energy through magnetic reconnection (Drenkhahn 2002) or current driven instabilities 
(Lyutikov \& Blandford 2003; Giannios \& Spruit 2006) can directly power the prompt emission.     

If the dissipated energy leads to fast particles, then the synchrotron
or synchrotron self Compton mechanism appears as a natural one for the prompt
emission. On the other hand, the large number of bursts with low energy
slopes steeper than the synchrotron model predictions (e.g. Crider et al. 1997;
Frontera et al. 2000; Girlanda et al. 2003), have triggered alternative suggestions
for the origin of the GRB emission. These include saturated Comptonization (Liang 1997),
Comptonization by thermal electrons (Ghisellini \& Celotti 1999) and the
photospheric emission of the flow (M\'esz\'aros \& Rees 2000; Ryde 2004). 

Both fireball and magnetic models predict some degree of photospheric emission
that comes from the region where the flow becomes optically thin to Thomson scattering.
If no energy dissipation takes place in the photospheric region, this emission is expected 
to be quasi-thermal and its peak should be clearly observed in the case
of fireball models (Daigne \& Mochkovitch 2002); which is not the case. On the other hand,
if there is substantial energy dissipation (through internal shocks, magnetic
reconnection or MHD instabilities) close to the photospheric region, the emitted spectrum 
can strongly deviate from a black body  (M\'esz\'aros \& Rees 2000; Pe'er et al. 2005, 2006; 
Ramirez-Ruiz 2005).

In this work, I study the spectrum of the photospheric emission in dissipative GRB models.
I use the magnetic reconnection model (Drenkhahn 2002; Drenkhahn \& Spruit 2002)
since it makes robust predictions for the characteristics of the flow and the rate 
of magnetic energy dissipation at different radii and study in detail the radiative transfer 
in this model. Similar considerations can be applied to other dissipative 
models (see also Pe'er et al. 2006). 

The study shows that while deep inside the flow radiation 
and matter are in approximate thermodynamic equilibrium, direct heating of the electrons
through energy dissipation leads to increase of their temperature as a function radius 
at radii where the flow is still thick to Compton scattering (about a factor of ten below the 
photospheric radius). 

A Monte Carlo code has been implemented to study the effect of Compton 
scattering at the photospheric region. The electron 
temperature is determined by balancing the heating by dissipation with the
cooling by Comptonization. 
The calculations show that the flow develops a hot photosphere with 
comoving electron temperatures of the order of $\sim 30$ keV (and larger further out)
for a large range of luminosities and baryon loadings of the flow. 
Inverse Compton scattering of the underlying thermal radiation leads
to a Band-like spectrum which peaks at $\sim 1$ MeV (in the central engine frame)
followed by a nearly flat high energy spectrum (with $E\cdot f(E)\sim E^0$) that 
extends up to a few hundred MeV.  

In the next Section, I discuss various dissipative models for GRB outflows and 
review the main features of the magnetic reconnection model. 
Analytical estimates for the radiative transfer and the electron temperature
in the photospheric region are given in Sect.~3. The description of the
Monte Carlo code and the numerical results are presented in Sect.~4. The last two Sections
contain the discussion and the conclusions.    

\section{Dissipation close to the photosphere}
\label{model}

The fireball model for GRB outflows (Paczy\'nski 1986; Goodman 1986) predicts 
the emission of a photospheric component from the region where radiation and matter decouple. 
This decoupling takes place close to the Thomson photosphere of the flow. 
The photospheric radiation can carry away most of the luminosity of the flow
(if the latter is still in its acceleration phase when the decoupling
takes place) or a smaller fraction $\sim (r_{\rm{s}}/r_{\rm{ph}})^{2/3}$ of it
depending on the baryon loading of the flow. Here $r_{\rm{s}}$ stands for the 
saturation radius where the flow reaches its terminal Lorentz factor and
$r_{\rm{ph}}$ is the photospheric radius. In the absence of energy dissipation 
around $r_{\rm{ph}}$, the photospheric component is expected to be quasi-thermal.
Although a small fraction of GRBs do show thermal emission (Ryde 2004),
the typical prompt GRB spectrum cannot be fitted by a black body. 

Internal shocks (Rees \& M\'esz\'aros 1994) in an unsteady flow are usually invoked
to dissipate a fraction of the bulk kinetic energy of the flow into fast
particles and random magnetic fields and to power the prompt emission. For a range
of the parameter space of the internal shock model, these shocks can take place below
the Thomson photosphere, influencing the strength and the spectrum of the photospheric
component (Rees \& M\'esz\'aros 2005; Pe'er et al. 2005, 2006). 

In the case of Poynting-flux dominated flows, both acceleration of the flow and 
emission of radiation can take place through dissipation of magnetic energy. 
Magnetic energy can be dissipated directly through reconnection in a 
flow where the magnetic field changes polarity on small scales (AC flow; Drenkhahn 2002;
Drenkhahn \& Spruit 2002) or via MHD instabilities in an axisymmetric flow 
(Giannios \& Spruit 2006). The magnetic dissipation and acceleration in these models is
gradual and takes place over several decades in radius that typically include the photospheric
radius. The photospheric component can be quite strong,
of the order of $\simless 20$\% of the luminosity of the flow. 

Here, I focus on the observational appearance of the photospheric component in an ``AC'' flow.    
The basic features of the dynamics of the flow are reviewed in the next Section, while
the detailed dynamical calculations are presented in Drenkhahn (2002) and
Drenkhahn \& Spruit (2002).

\subsection{The reconnection or ``AC'' model}

An important physical quantity of the flow in the ``AC'' model is the ratio 
$\sigma_0$ of the Poynting flux to kinetic energy flux at the Alfv\'en point $r_0$.
This is a quantity that plays similar role to the ``baryon loading'' parameter that
appears in fireball models. the ratio  $\sigma_0$ determines the 
terminal bulk Lorentz factor of the flow $\Gamma_\infty$ which is of the order
of $\sim \sigma_0^{3/2}$. The flow must start Poynting-flux      
dominated with $\sigma_0\simmore 30$ for it to be accelerated to ultrarelativistic
speeds with $\Gamma_\infty \simmore 100$. 

In an ``AC'' model, such as produced by an inclined rotator (Coroniti 1990),
the magnetic field in the flow changes 
polarity on small scale $\lambda$ of the order of the light cylinder 
in the central engine frame (i.e. $\lambda \simeq 2\pi c/\Omega$, where
$\Omega$ is the angular frequency of the rotator).
In the flow, the energy density of the magnetic field is larger than
the rest mass energy density and the reconnection speed is increased by 
the relativistic kinematics to subrelativistic speeds (Lyutikov \& Uzdensky 2003; Lyubarsky 2005). 
Magnetic reconnection is modeled to proceed at a fraction 
$\varepsilon \sim 0.1$ of the Alfv\'en speed (which is essentially the speed of light for 
a magnetically dominated flow). Because of dissipation of the magnetic energy, about half of the 
Poynting flux converts into kinetic flux (i.e. acceleration of the flow) and the other half into
internal energy of the flow. The dissipation is gradual and takes 
place up to the ``saturation radius'' $r_{\rm{s}}$ where reconnection
stops and the flow achieves its terminal Lorentz factor.

Under the assumption of 1-D, steady flow, the relativistic MHD equations
can be solved analytically for radii $r_0\ll r\ll r_{\rm s}$, yielding the following 
self similar scalings for the comoving number density and the comoving magnetic field strength (Drenkhahn 2002)
\be
n'=\frac{1.5\cdot 10^{17}}{r_{11}^{7/3}}\frac{L_{52}}{(\varepsilon \Omega)_3^{1/3}\sigma_{0,2}^2}
\qquad\rm{cm^{-3}},
\label{density}
\ee
\be
B'=\frac{1.4\cdot 10^8}{r_{11}^{4/3}}\frac{L_{52}^{1/2}}{(\varepsilon \Omega)_3^{1/3}\sigma_{0,2}^{1/2}}
\qquad\rm{Gauss},
\label{Bfield}
\ee
respectively. The notation $A=10^xA_x$ is used and the ``reference values'' of the
model parameters are $\sigma_0=100$, $\varepsilon=0.1$, $\Omega=10^4$ rad s$^{-1}$, $L=10^{52}$ 
erg s$^{-1}$ sterad$^{-1}$; very close to those used in previous studies of the model (see Drenkhahn 2002; 
Drenkhahn \& Spruit 2002; Giannios \& Spruit 2005).  

The model predicts gradual acceleration of the flow with $\Gamma\sim r^{1/3}$ in the 
regime $r_0\ll r\ll r_{\rm s}$, while no further acceleration takes place above the
saturation radius. The bulk Lorentz factor of the flow is, thus, approximately given 
by the expression
\begin{eqnarray}
\Gamma&=&\Gamma_\infty\left(\frac{r}{r_{\rm s}}\right)^{1/3}=148r_{11}^{1/3}(\varepsilon \Omega)_3^{1/3}
\sigma_{0,2}^{1/2}, \quad\rm{for}\quad r<r_s \nonumber\\
&&\\
\Gamma&=&\Gamma_\infty=\sigma_0^{3/2}, \quad\rm{for}\quad r\ge r_s. \nonumber
\label{gamma}
\end{eqnarray}
The saturation radius is given by 
\be
r_{{\rm s,11}}=\frac{\pi c \Gamma_\infty^2}{3 \varepsilon\Omega}=
310\frac{\sigma_{0,2}^3}{(\varepsilon \Omega)_3}.
\label{rsatur}
\ee

Another characteristic radius of the flow is the Thomson photosphere.
The optical depth between two points in the flow can be found by integrating
the expression ${\rm d}\tau=\Gamma(1-\beta\cos \theta)n' \sigma_{\rm T} {\rm d}s$ (Abramowicz
et al. 1991), where $\beta=v/c$ and $\theta$ the angle of a photon
path with respect to the radial direction. Integrating the previous expression
from $r$ to $\infty$ for a radially moving photon, one gets the 
characteristic Thomson optical depth as a function of radius 
\be
\tau=\frac{20}{r_{11}^{5/3}}\frac{L_{52}}{(\varepsilon \Omega)_3^{2/3}
\sigma_{0,2}^{5/2}}.
\label{tau}
\ee
The location of the photosphere is found by setting $\tau=1$ and
solving for $r$
\be
r_{\rm{ph,11}}=6\frac{L_{52}^{3/5}}{(\varepsilon \Omega)_3^{2/5}
\sigma_{0,2}^{3/2}}.
\label{rphot}
\ee

The location of the photospheric radius with respect to the saturation one
determines the nature of the emitted spectra expected from the flow.
If, for example, $r_{\rm{ph}}\gg r_s$, all the energy dissipation takes place 
in optically thick conditions and the radiation is efficiently thermalized.
On the other hand, if $r_{\rm s}\simmore r_{\rm{ph}}$, energy dissipation at 
moderate and low optical depths leads to a photospheric emission that 
has a highly non-thermal appearance that can be directly related 
with the characteristic GRB emission. These points become clear 
in the next sections where the radiative transfer around the region
of the photosphere is studied with analytical and numerical tools.    

\section{Analytical estimates}

Energy dissipation in an ``AC'' flow
is gradual and takes place over many decades of radii up to the
saturation radius $r_{\rm s}$. Half of the dissipated Poynting flux 
serves to accelerate the flow, while the other half is released
as internal energy in the flow. The rate of energy density release in a comoving
frame can be found by the following considerations. The time scale
over which the magnetic field decays is that of advection of 
magnetic field of opposite polarity to the reconnection area.
The reconnection speed is $v_{r}=\varepsilon v_{\rm A} \simeq \varepsilon c$,
while the magnetic field changes polarity over a length scale
$\lambda'=2\pi \Gamma c/(\varepsilon\Omega)$ measured in a frame comoving with the
flow. The decay timescale for the magnetic field, therefore, is
\be
t_{{\rm dec}}=\frac{\lambda'}{v_r}=\frac{2\pi\Gamma}{\varepsilon\Omega}.
\ee  
Using the last expression and eqs. (\ref{Bfield}) and (\ref{gamma}), 
the rate of magnetic energy density dissipation in the comoving frame is 
\be
P_{\rm{diss}}=\frac{B'^2/8\pi}{t_{{\rm dec}}/2}=\frac{1.5\cdot10^{15}}{r_{11}^3}
\frac{L_{52}}{\sigma_{0,2}^{3/2}} \quad \rm{erg cm^{-3} s^{-1}}.
\label{Pdiss}
\ee  
The fate of the released energy is rather uncertain. An interesting possibility
is that dissipation leads to MHD turbulence where particle acceleration can
take place by scattering of photons by Alfv\'en waves at the $\tau \sim 1$
region of the flow (Thompson 1994). On the other hand, the magnetic energy can directly 
be dissipated to the particles in the flow, most likely to the electrons due to their 
higher mobility. Here I assume that a fraction $f_{\rm e}$ of order unity of the energy 
is heating the electrons of the flow. The electrons are assumed to have a thermal
distribution. Deep in the flow, the relaxation timescale of the electrons due to
Coulomb collisions is fast enough to ensure thermalization. I return to this
issue in Sect.~4.3.

Deep inside the flow the released energy is efficiently thermalized and shared
between particles and radiation. Assuming complete thermalization, integration
of the energy released at different radii in the flow, taking into account
adiabatic cooling, leads to the following expression for the comoving
temperature of the flow (Giannios \& Spruit 2005)
\be
T_{\rm{th}}=\frac{0.7}{r_{11}^{7/12}}\frac{L_{52}^{1/4}}{(\varepsilon \Omega)_3^{1/12} 
\sigma_{0,2}^{1/2}} \quad \rm{keV}.
\label{Tth}
\ee   
Comoving temperatures of a few hundred eV are typical at the region where the
optical depth of the flow is of order of unity (see the last expression and 
eq. (\ref{rphot})). At these temperatures the Compton 
scattering cross section dominates over free-free absorption so that
deviation from thermal spectra is expected. It turns out, however, that
the assumption of equilibrium of radiation and particles breaks down
deeper in the flow and that the ``photospheric'' spectra are
highly non-thermal for typical GRB parameters.

The last point becomes clear if one looks at the energy balance of the electrons
in the flow. The electrons are heated by magnetic energy dissipation
and cool mainly through radiative cooling (the adiabatic cooling and the 
energy exchange with the protons are much slower). An obvious 
candidate for radiative cooling of the electrons in a strongly magnetized 
flow is synchrotron cooling since the energy density of the magnetic field
is larger than that of radiation 
\be
\frac{U_{\rm B}}{U_r}=\frac{B'^2/8\pi}{aT_{\rm{th}}^4}=23\frac{\sigma_{0,2}}
{r_{11}^{1/3}(\varepsilon \Omega)_3^{1/3}}.
\label{ratio}
\ee      
This suggests dominance of synchrotron over Compton cooling {\it if}
the flow is optically thin to synchrotron emission. It turns out that
this is {\it not} the case in the photospheric region under consideration
and that synchrotron radiation is strongly self-absorbed (I 
return to this issue in the end of this Section). The dominant
cooling process for the electrons is, thus, Compton cooling. 

The Compton cooling rate for the electrons is given by the expression
(e.g. Rybicki \& Lightman 1979)
\be
P_{\rm {Comp}}=4n'\Theta_{\rm e}c\sigma_{\rm T}U_r,
\label{Pcomp}
\ee
where $\Theta_{\rm e}=K_BT_{\rm e}/(m_{\rm e}c^2)$ and $\sigma_{\rm T}$ is the Thomson scattering 
cross section. Equating the heating rate (given by eq.~(\ref{Pdiss})) to the cooling
rate of the electrons, I solve for the electron temperature
\be
T_{\rm e}=2.0 r_{11}^{5/3}f_{{\rm e,1}}\frac{(\varepsilon \Omega)_3^{2/3}
\sigma_{0,2}^{5/2}}{L_{52}}\quad \rm{keV},
\label{Te}
\ee
where $f_{\rm{e,}1}$ stands for $f_{\rm e}=1$.
So the electron temperature increases as a function of radius
and equilibrium of the electrons and radiation is reached
deep in the flow at the radius where $T_{\rm{th}}=T_{\rm{e}}$. The last equation 
defines a characteristic radius which I call the {\it equilibrium radius} 
\be
r_{\rm{eq,11}}=0.6\frac{L_{52}^{5/9}}{f_{{\rm e,1}}^{4/9}(\varepsilon \Omega)_3^{1/3}\sigma_{0,2}^{4/3}},
\label{req}
\ee  
where eqs. (\ref{Tth}) and (\ref{Te}) have been used to derive the last expression.
Note that equilibrium of matter and radiation is achieved at a radius
that is factor of ten shorter than that of the Thomson photosphere, or at an
optical depth (combining eqs. (\ref{tau}) and (\ref{req}))
\be
\tau_{\rm{eq}}=46\frac{f_{{\rm e,1}}^{20/27}L_{52}^{2/27}}{(\varepsilon \Omega)_3^{1/9}\sigma_{0,2}^{5/18}}
\label{taueq}
\ee
which depends weakly on the model parameters and is much larger than unity.

At radii $r>r_{\rm{eq}}$, the electron temperature is higher than that of the radiation
field and upscattering of the photons takes place. Because of the increase of the
electron temperature with radius, this upscattering becomes more efficient 
close to the location of the photosphere. The electron temperature there
 can be found by combining eqs. (\ref{rphot}) and
(\ref{Te}) which give $T_{{\rm e,ph}}=40 f_{{\rm e,1}}$ keV {\it independently} of  
the parameters of the model except to the fraction $f_{\rm{e}}$.
For temperatures $T_{\rm e}\simmore 40$ keV in the $\tau_{\rm T}\simless 1$ region, the
characteristic signature of unsaturated Comptonization is expected to lead
to spectra with hard non-thermal appearance. 
 
The previous estimates are based on the assumption that synchrotron emission
in the flow is strongly self-absorbed and therefore the electrons cool
mainly through Compton upscattering the radiation field. I can now check the 
validity of this assumption by estimating the synchrotron (and free-free) 
absorption optical depths in the flow and the characteristic turn-over frequency 
below which radiation becomes optically thick. The absorption optical depth for 
a photon traveling from radius $r$ to $2r$ is given by the expression
$\tau_\nu=\alpha_\nu r/\Gamma$, where $\alpha_\nu$ is the absorption
coefficient as measured by the comoving observer. Note that I have not
used the ``effective'' optical depth even though the inner parts of the 
flow are Thomson thick. This is justified by the relativistic nature of the
problem. A photon that is emitted at a radius $r$ that corresponds to an optical depth
$\tau\gg 1$, {\it does not} undergo $\sim \tau^2$ scatterings until it escapes as it would
be expected from random walk arguments. It suffers
$\sim \tau$ scatterings since it preferentially scatters along the radial
direction (with an typical angle $\theta\sim 1/\Gamma\ll 1$ with respect to the radial
direction) in the central engine frame. 

For a thermal plasma and for characteristic comoving photon energy $h\nu\ll k_{\rm B}T_{\rm e}$,
the absorption coefficient is related to the emission $j_\nu$ through the
well known expression $\alpha_\nu=j_\nu/2\nu^2m_{\rm e}\Theta_{\rm e}$. For the mildly relativistic
plasma under consideration, the synchrotron emission is well approximated by eq.~(13)
of Wardzi\'nski \& Zdziarski (2000; see also Petrosian 1981; Petrosian 
\& McTiernan 1983). Using their expressions for synchrotron emission, I have
calculated the turnover frequency $\nu_t$, defined by the expression $\tau_{\nu_t}=1$
at different radii in the flow and have verified that $\nu_t\gg \nu_c$ (where $\nu_c=eB'/2\pi 
m_{\rm e} c$ is the cyclotron frequency) for a large range of the model parameters and up to radii
that correspond to low optical depths $\sim 0.01$ (or corresponding radii given by 
eq.~\ref{tau}); sufficient for this study. Below $\nu_t$ the synchrotron emission is
strongly self-absorbed, resulting in synchrotron cooling rates in the plasma orders
of magnitude less than the Compton cooling rates.

As the analytic estimate (\ref{Te}) shows, the electron temperature increases
rather fast and can upscatter the photon field causing deviations
from thermal spectra at a radius of about a factor of ten below the Thomson 
photosphere. On the other hand, this estimate has its limitations since it 
gives the electron temperature by balancing the electron heating of the
electrons to the Compton cooling rate assuming thermal distribution of photons.
A more accurate calculation involves a self consistent determination of the
electron temperature simultaneously with the actual distribution of the photon field. 
This calculation is the topic of the next Section.

\section{Numerical Study}

To study the emergent spectrum and the electron temperature above the 
``equilibrium'' radius in detail, a Monte Carlo code has been developed. 
It simulates the Compton scattering in a flow with density and bulk
Lorentz factor given by eqs. (\ref{density}) and (\ref{gamma}) 
respectively, by following the scattering random walk of a large number of photons
(Pozdniakov et al. 1983). The special relativistic effects related to both the bulk
motion and the scattering cross section of photons in the flow are taken into account. 
The inner boundary of the computational domain is taken at the last radius where the 
electrons are in equilibrium with the
radiation field, i.e. the equilibrium radius defined by eq. (\ref{req}).
At this radius a black body photon spectrum is injected with comoving temperature given by 
eq. (\ref{Tth}) evaluated at $r_{\rm eq}$ (given by eq. (\ref{req}))
\be
T_{{\rm eq}}=0.9\frac{f_{{\rm e,1}}^{7/27}(\varepsilon \Omega)_3^{1/9}\sigma_{0,2}^{5/18}}{L_{52}^{2/27}}
\quad \rm{keV}.
\label{Teq}
\ee
The outer boundary of the computational domain is at $r_{{\rm out}}$ where the optical
depth becomes small (the value $\tau=0.1$ is used).  

Obviously the result of the Comptonization depends critically on the electron 
temperature which, in turn, depends on the Compton cooling rate. The 
accurate determination of $T_{\rm e}(r)$ demands an iterative method where 
the flow is divided in a large number of spherical shells (of the order of $10^2$) and 
an initial guess for $T_{\rm e}(r)$ (e.g. the analytic estimate of eq. (\ref{Te})) is given. 
Then $T_{\rm e}$ is varied in the different shells until balance of the heating and cooling rates
is achieved at all radii. This method is, however, rather computationally demanding. Fortunately, 
it turns out that one can get very accurate results assuming a power-law dependence of the electron
temperature with radius of the form
\be
T_{\rm e}=T_0\left(\frac{r}{r_{{\rm e}q}}\right)^s,
\label{Tenum}
\ee
where $T_0$ and $s$ are two free parameters to iterate so that heating 
approximately balances cooling at every radius. The results of such iteration 
and the emitted photospheric spectra and their energetics for various values
of the parameters of the model are the subject of the next Section. 

\subsection{Results} 

The first step toward calculating the photospheric spectra 
is to determine the electron temperature parameters $T_0$ and $s$ for which 
Compton cooling balances the heating rate of the electrons at {\it every} radius in the flow. 
A first guess for  $T_0$ and $s$ comes from the analytic estimate (\ref{Te}).

\begin{figure}
\resizebox{\hsize}{!}{\includegraphics[angle=270]{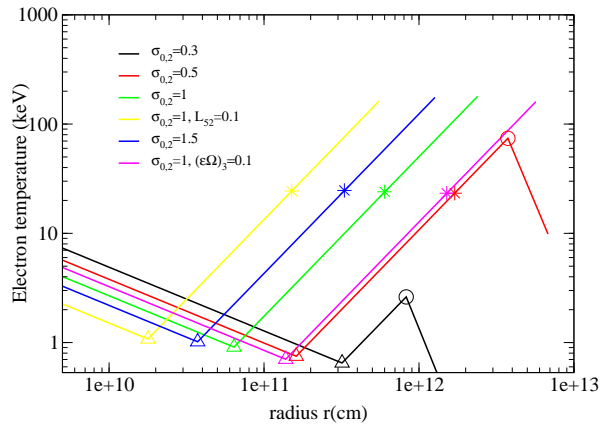}}
\caption[]
{The comoving electron temperature as a function of radius from the numerically
determined energy balance with the radiation field. The baryon loading
of the flow is given in the legend. The parameters that are not given are 
taken to have their reference values (see Sect. 2.1). For small radii radiation and the 
electrons are in thermal equilibrium. Triangles mark the radius where radiation and
electron temperature get out of equilibrium,  asterisks mark the photospheric radius
and circles mark the saturation radius. Note that the electron temperature
at the photosphere is $\sim 25$ keV for all the parameter range, except from the
high baryon loading (or low $\sigma_0$) cases.   
 
\label{fig1}}
\end{figure}

In figure 1, I plot the resulting electron temperature as a function of 
radius for different values of the parameters. At small radii, radiation and
matter are in thermal equilibrium. Above the equilibrium 
radius introduced in eq. (\ref{req}), the electron temperature increases following the
expression (\ref{Tenum}) for the numerically computed parameters $T_0$ and
$s$ which are close to the analytic estimate $T_{\rm{eq}}$ and $5/3$ respectively.
The asterisks in Fig.~1 show the location of the Thomson photosphere.
Note that the temperature at the location of the photosphere is rather
independent on the parameters of the model and about $\sim 25$ keV,
slightly lower than the analytical estimate (\ref{Te}). 
For the high $\sigma_0$ cases, the electron temperature keeps increasing 
until the outer boundary of the simulation. For moderate and high $\sigma_0$,
the temperature is $\sim 200$ keV at this outer boundary. The 
very similar temperatures at the same optical depths in the flow lead 
to similar $y$-Comptonization parameter and, therefore, similar emitted spectra 
for a large range of the parameters of the model. This point becomes clear
further down in this Section where the computed spectra are presented.  

Note that in the low $\sigma_0$
(or large baryon loading) cases, there is a point above which the
electron temperature  drops again as a function of radius. 
This point is the saturation radius $r_s$ (defined by eq.~(\ref{rsatur}) and marked
by circles in Fig.~1). Above this radius no significant dissipation of magnetic energy 
takes place and the heating rate of the electrons drops quickly. The region $r>r_{\rm s}$  
does not contribute much to the emergent spectra (except from some degree of adiabatic cooling)
since the electrons become essentially cold.      

\begin{figure}
\resizebox{\hsize}{!}{\includegraphics[angle=270]{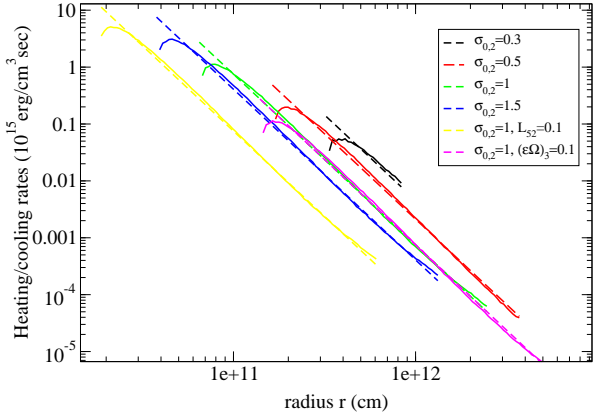}}
\caption[]
{The heating and cooling rates, dashed and solid lines respectively, as a function of 
radius for different parameters of the flow given in the legend. The rest of the parameters are 
assumed to have their reference values (see Sect.~2.1). The cooling and heating rates balance each
other well, except from a narrow region close to the ``equilibrium'' radius in
the inner boundary of the computation domain. This illustrates that a power-law is a good
approximation for the dependence of the electron temperature with radius.

\label{fig2}}
\end{figure}

In figure 2, I show the cooling and heating rate as a function of radius
for the numerically iterated values of $T_0$ and $s$.
Note that the power law modeling for the
radial dependence of the electron temperature (see eq. (\ref{Tenum}))
leads to a very good  balance  except
in the very inner region of the computed domain. So the 
modeling of $T_{\rm e}$ appears to be accurate enough for the purpose
of this calculation. Obviously, a smoother fitting function could be
used around the equilibrium radius, which 
would give a better description in this region. However, since the
transition region is rather narrow, only a small fraction of the energy
is dissipated there. This simplification is not expected to 
introduce large errors to the calculations and more detailed modeling 
of the transition has not been pursued.   

The emerging spectra from our calculations are given in Fig.~3 in $E\cdot f(E)$ 
representation for different values of the parameters of the model.
The spectra are plotted in the central engine frame with arbitrary 
normalization (the energetics are discussed in the next paragraphs).
The peak of the spectrum is close to $~1$ MeV for a large region of 
the parameter space which is relevant for GRB flows.  The 
clustering of the peak around 1 MeV can be understood by the following 
considerations. The temperature of the photons at the equilibrium radius
in the central engine frame is
\be
T_{\rm eq}^{\rm ce}=\frac{4}{3}\Gamma(r_{\rm {eq}})T_{\rm eq}=150
f_{{\rm e,1}}^{1/9}L_{52}^{1/9}(\varepsilon \Omega)_3^{1/3}\sigma_{0,2}^{1/3} \quad \rm{keV},
\label{Teqce}
\ee 
where in the last step the expressions (\ref{Teq}), (\ref{gamma}) and (\ref{req}) have been used
\footnote{Note that this expression for the photon
temperature does not apply for the very low $\sigma_0\simless 25-30$ cases for which dissipation does not 
proceed up to $r_{\rm eq}$ which is the condition for the expressions (\ref{Tth}) and, therefore, (\ref{Teq})
and (\ref{Teqce}) to be applicable.}.
The photon spectrum at $r_{\rm{eq}}$ peaks at $\sim 4 T_{\rm eq}^{\rm ce}\sim 600$ keV 
(in a $E\cdot f(E)$ representation) for the reference values of the parameters and depends very 
weakly on the flow parameters (see eq. (\ref{Teqce})). The inverse Compton scattering that takes 
place above the equilibrium radius
leads to a moderate increase of the energy of the peak and, most important, to a 
power-law high energy emission with photon number index $\sim -2.3$
that extends up to a few hundred MeV. This high energy part of the spectrum
is the result of unsaturated Comptonization taking place close to the 
Thomson photosphere and appears for flows with $\sigma_0\simmore 50$. 

Both the peak and the high energy slope lie well within the observed range of the typical
GRB spectrum, confirming that the photospheric components may be responsible for the prompt 
GRB emission. The low energy
slope of the spectrum depends on the energy range over which one attempts to fit
it. Far from the peak, its slope is about $f(E)\propto E^2$ (i.e. with photon number index
$\alpha\sim 1$). When fitted in the BATSE range (assuming a burst at z=2) with the Band 
spectrum (Band et al. 1993), typical values of $\alpha\sim -0.3$ are found for the
spectral slope below the peak frequency. These values are compatible with those
measured in hard bursts and cannot be explained with the simplest synchrotron models. 

\begin{figure}
\resizebox{\hsize}{!}{\includegraphics[angle=270]{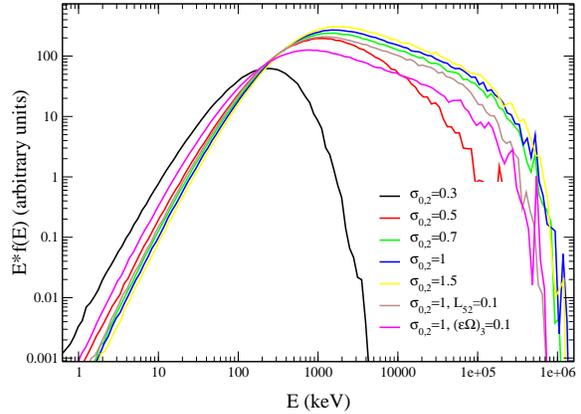}}
\caption[]
{The resulting $E\cdot f(E)$ spectrum for different values of the parameters 
of the flow (photon energies in the central engine frame). 
The moderate and high $\sigma_0$ cases exhibit similar spectra with a peak 
that clusters at $\sim 1$ MeV and a flat high energy tail that is a result
of unsaturated Comptonization near the photospheric radius.
For low $\sigma_0$ dissipation stops below the photosphere and the
resulting spectrum is quasi-thermal. 
\label{fig3}}
\end{figure}

In rather low $\sigma_0\simless 50$ flows, the emerging spectrum has much weaker
emission above its peak. In these cases the magnetic dissipation stops
close to, or even below, the Thomson photosphere and there is
only weak Compton upscattering taking place in the photospheric
region. Quasi-thermal emission has been observed in a fraction of
GRBs (Ryde 2004; 2005) and they may be a result of low $\sigma_0$
(i.e. high baryon loading).   For even smaller values of  $\sigma_0\simless 25-30$, 
the magnetic dissipation stops bellow the equilibrium radius and the photon field 
suffers substantial adiabatic cooling before it decouples from matter. This results in 
quasi-thermal photospheric spectrum that peaks at tens of keV or lower. 

The X-ray flashes are events that have spectral properties very similar to these of the 
classical GRBs but with spectral peak below $\sim 50$ keV and are
believed to belong to the same family with GRBs (e.g. Barraud et al. 2003).
It is tempting to identify the X-ray flashes with these very low $\sigma_0$
flows. On the other hand, the spectrum of the X-ray flashes is similar
to that of the classical GRBs which makes this identification unlikely. 
It appears more natural, in the context of this model, that X-ray flashes are
typical GRB flows viewed off-axis (and therefore with low $E_{\rm peak}$). 
Although the issue on the nature of the X-ray flashes is not settled, afterglow modeling 
seems to support this interpretation (e.g. Granot et al. 2005).     

\subsection{Efficiency of the process}

A question quite relevant to the observational relevance of the 
photospheric component is its strength. A convenient quantity
to measure the strength of this component is the photospheric
efficiency $e_{\rm{ph}}$ defined here as the ratio of the photospheric
luminosity to the total luminosity of the flow. In Fig.~4,
I plot the $e_{\rm{ph}}$ for various $\sigma_0$. For large 
$\sigma_0$, the efficiency is rather low $\simless 10$\%
since most of the magnetic energy is dissipated further out in the
flow. For moderate values of $\sigma_{0,2}\sim 0.5$, the photospheric
component becomes stronger with $e_{\rm{ph}}\simmore 15$\%, while 
its strength is reduced again for low $\sigma_0$. In this case,
the dissipation stops when the flow is still Thomson thick and
radiation is cooled adiabatically before it decouples from matter.
These numbers correspond to $L_{52}=1$, $(\varepsilon \Omega)_3=1$. 
Higher flow luminosities and values of $\varepsilon \Omega$ lead to
higher efficiencies and vice versa (see Fig.~4). 

In Fig.~4, the ``photospheric efficiency'' in the
BATSE range $e_{ph,B}$ is also plotted. The BATSE range is taken to be
(30-2000) keV which correspond to the (30-2000)$\times$(1+z) keV energy 
range in the central engine frame. For the redshift, I  take z=2.       
In most of the parameter space explored, more than $\sim$60\% of the 
photospheric emission is in the BATSE range; resulting in $e_{\rm{ph,B}}$ 
close to $e_{\rm{ph}}$ (see Fig.~4).     

\begin{figure}
\resizebox{\hsize}{!}{\includegraphics[angle=270]{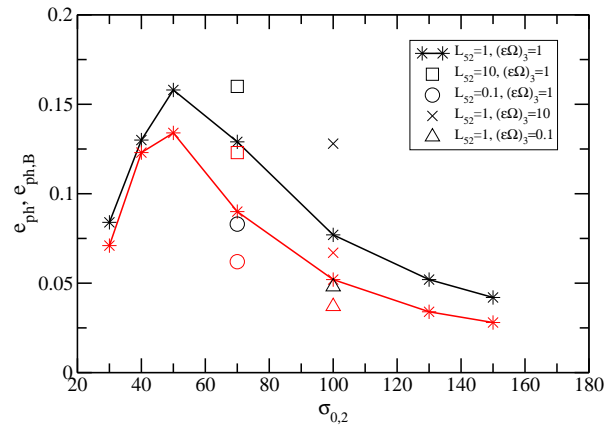}}
\caption[]
{The ratio of the photospheric luminosity in the whole energy range
(black symbols) and in the BATSE energy range (red symbols) to
the flow luminosity. Overall the photospheric radiative efficiency 
varies in the range a few to 20\%. Higher flow luminosities and values of 
$\varepsilon \Omega$ increase the radiative efficiency and vice versa.

\label{fig4}}
\end{figure}

In the high $\sigma_0$ cases, additional radiation is expected to
come from outside the computed domain since magnetic dissipation
continues up to $r_{\rm{s}}\gg r_{\rm{ph}}$. In this Thomson thin region,
various of the assumptions underlying our calculations may not
hold anymore. I have assumed a thermal distribution of the 
electrons which can no longer be justified outside the photosphere
(see next Section). Furthermore, the electron temperature increases
rapidly with radius. At some point it approaches  that of the rest mass of 
electrons and pair creation should also be taken
into account. Last, the non-thermal electron distributions
are expected to lead to efficient synchrotron emission and synchrotron
cooling cannot be neglected anymore. The problem of radiative transfer in this
Thomson thin region depends critically on the particle acceleration process
 in the regions where magnetic reconnection takes place.
Some estimates on the processes at work in this region can be found in 
the Giannios \& Spruit (2005).

I discuss the issues of the thermalization of the electron distribution 
and of pair creation more quantitatively in the next subsections.

\subsection{Thermalization of the electrons}

In our study, I assumed a thermal distribution for the electrons. Particle acceleration
in the current sheet in the reconnection regions can lead to highly non-thermal
particle distributions but, if the time scale of energy exchange through Coulomb
collisions is shorter than the cooling timescale, thermalization of the
electron distribution is achieved. The relaxation timescale for the electrons
is given by (Spitzer 1967; Stepney 1983)
\be
t_{\rm{ee}}=\frac{4\sqrt{\pi}}{\ln \Lambda}\frac{\Theta_{\rm e}^{3/2}}{n'\sigma_{\rm T}c}=
4\cdot10^{-8}r_{11}^{29/6}\frac{f_{{\rm e,1}}^{3/2}(\varepsilon \Omega)_3^{4/3}
\sigma_{0,2}^{23/4}}{L_{52}^{5/2}}\quad \rm{s},
\label{tee}
\ee
where $\ln \Lambda$ is the Coulomb logarithm. For the typical densities
and temperatures of the problem at hand the value $\ln \Lambda=13$ is used. 
I have also used the eq. (\ref{density})
and the estimate (\ref{Te}) for the electron temperature (the use of the
more accurate numerically calculated values for $T_{\rm e}$ leads to very similar results).

The cooling timescale of the electrons is 
\be
t_{\rm{cool}}=\frac{\frac{3}{2}n'k_{\rm B}T_{\rm e}}{P_{\rm{Comp}}}=5\cdot
10^{-7}r_{11}^{7/3}\frac{(\varepsilon \Omega)_3^{1/3}
\sigma_{0,2}^2}{L_{52}} \quad \rm{s},
\label{tcool}
\ee 
where the eqs.~(\ref{Pcomp}), (\ref{Te}), (\ref{Tth}) and (\ref{density})
are used in the last step. So, at radius of about $10^{11}$ cm the
relaxation timescale due to Coulomb collisions is shorter than the
cooling timescale. However, the former has steeper radial dependence and the two
timescales become equal at a radius $r*$
\be
r*_{11}=3\frac{L_{52}^{3/5}}{f_{{\rm e,1}}^{2/5}(\varepsilon \Omega)_3^{2/5}
\sigma_{0,2}^{3/2}}.
\label{rstar}
\ee
   
This radius is a factor of 2 inside the photospheric radius for
$f_{e}\simeq 1$ (compare the last expression with eq.~(\ref{rphot})).
Thermalization of the electron distribution is therefore a fair
approximation inside the photosphere.   
Above the location of the photosphere, deviations from thermal distribution
are expected. The actual electron distribution, however, is still
expected (from energetic considerations) to peak at energies where 
heating and cooling rates balance each other. 

\subsection{Pair Creation}      

In both the analytical estimates and the numerical investigation, I have neglected
pair creation. This is justified close to the equilibrium radius
where the photon field is thermal with comoving temperature of about
1 keV; i.e. much below the pair creation threshold (see minimum
of the curves in Fig.~1). However, the electron temperature increases further
out in the flow, reaching values of the order of 200 keV near the end
of our computational domain at $\tau =0.1$. A fraction of the photons are expected to 
be upscattered above the pair creation threshold in the comoving frame and 
lead to pair creation. To estimate how efficiently pair creation takes place, 
one can define the ``comoving compactness'' $l'=n_{\gamma}\sigma_{\rm T}r/\Gamma$, where 
$n_{\gamma}$ is the number density of photons that exceed the rest mass energy of the
electrons (in the comoving frame). It can be shown (Lightman 1982; Svensson 1982;
Pe'er \& Waxman 2004) that if $l'$ is much larger than unity, extensive pair creation takes place 
in the flow and the number of pairs can exceed that of the electrons related to baryons.
On the other hand, if $l'\simless 1$ pair creation is not substantial. 

I have calculated $n_{\gamma}$ directly from our Monte Carlo simulations
at different radii and for different values of the parameters of the 
model. As expected, $l'$ is found to increase with radius and to
reach values up to $l'\simless 0.2$ for a large
parameter space investigated: $0.3\le\sigma_{0,2}\le 1.5$,
$0.01\le L_{52}\le10$, $0.1\le (\varepsilon\Omega)_3\le 10$, $0\le 
f_{\rm e}\le 1$. So the effect of pair creation is not expected to 
be important for parameters relevant for GRB outflows in the photospheric
region under consideration. 

This result can also be understood in view of Fig.~3. The high energy
part of the spectrum shows an exponential cutoff at $E_{c}\sim3\cdot 10^5$
keV. In the comoving frame, this feature appears at an energy $E_{c}/\Gamma
\sim m_{\rm e}c^2$. So only a small fraction (typically $10^{-3}$) of the 
photospheric luminosity is above the pair creation threshold, resulting
in rather low compactness to pair creation.

\section{Discussion}

In this work, I have investigated the appearance of the 
photospheric component in Poynting-flux dominated GRB outflows
in which the magnetic field changes polarity over small scales (AC model). 
An important characteristic of the model is that it predicts 
gradual dissipation of magnetic energy through reconnection
over many decades of radii including the region where
the flow has Thomson optical depths of order unity (Drenkhahn 2002; 
Drenkhahn \& Spruit 2002).

Here I have shown that, if a large fraction of the energy is
dissipated directly to the electrons, the electron temperature
increases rapidly at a wide region around the Thomson photosphere.
The electron temperature is self-consistently calculated at different
radii by balancing the heating and cooling rates. For this 
calculation both analytical estimates and Monte Carlo simulations 
are used. It is shown that inverse Compton scattering of the 
underlying thermal radiation leads to spectra that peak in the 1 MeV range 
(in the central engine frame) and have 
power-law high energy part for a wide range of the model parameters.
The high energy power law is a result is unsaturated Comptonization
that takes place at optical depths of order of unity. When fitted with the 
``Band'' spectrum (Band et al. 1993), the spectra
have low and high frequency spectral slopes and peak frequency in agreement
with observations. This ``photospheric''
component is a significant fraction (from $\sim 3$\% to more than $15$\%)
of the luminosity of the flow and may, therefore, be responsible for the prompt
GRB emission.

For high baryon loadings (low $\sigma_0$) in the flow, the energy dissipation
 stops below the Thomson photosphere and no spectral component
appears above the thermal peak. In this case, the photospheric emission
is quasi-thermal and may  be responsible for the appearance of 
a fraction of GRBs (Ryde 2004, 2005).   

Our calculation is limited to a region that extends up to a factor
of $\sim$ several above the photosphere (where the Thomson optical
depth drops to  $\tau\sim0.1$). However, models with sufficiently low
baryon loading (or equivalently high $\sigma_0$) predict energy 
dissipation that continues further out in the flow. The spectra 
expected by these outer parts of the flow have been to some extent
investigated by Giannios \& Spruit (2005). On the other hand, as long as
the issue of particle acceleration in the reconnection regions is poorly 
understood, definite predictions on the relative importance of radiative
mechanisms (e.g. Compton scattering, synchrotron emission) or the
emitted spectra are hard to be made about the Thomson thin part of the 
flow.

Our study was limited to the ``AC'' model because it makes robust
predictions concerning the dynamics of the flow and the rate of energy density 
dissipation. Furthermore, these predictions can take the form
of simple algebraic expressions that simplify our study of radiative
transfer in the flow. On the other hand, a large variety of models
have or may have significant energy dissipation in the photospheric region
that can lead to strong deviations from quasi-thermal spectra.
In the context of the internal shock model, for example, shocks 
can also take place close to the location of the photosphere, 
leading to modifications of the photospheric emission (M\'esz\'aros \&
Rees 2000; Ryde 2004; Rees \& M\'esz\'aros 2005; 
Pe'er et al. 2005). Another example of dissipative models is that
of a strongly magnetized flow with an axisymmetric magnetic field.
Such flow is subject to current driven instabilities (kink instability) 
that lead to gradual dissipation of magnetic energy and rather strong 
photospheric emission (Giannios \& Spruit 2006).

\section{Conclusions}

The standard fireball model for GRB flows predicts a rather strong photospheric
component that is emitted when radiation and matter decouple. A photospheric
component is also expected in magnetic GRB models (Drenkhahn \& Spruit 2002; 
Giannios \& Spruit 2006). In the absence of dissipative processes close to the photosphere, 
the photospheric component is expected to be quasi-thermal. In that case, a thermal peak
should be systematically observed in the prompt GRB spectra, which is not the case 
(Daigne \& Mochkovitch  2002).  
On the other hand, most of the GRB models for the prompt 
emission invoke energy dissipation at a large range of radii. 
If much energy dissipation takes place in the region of the photosphere, 
it can lead to large deviation of the photospheric component from purely thermal.
Such investigation has been made in the context of internal shock
and slow dissipation models by Pe'er et al. (2005).

In this work, I have focused on the photospheric emission
expected from a strongly magnetized outflow in which the magnetic field changes 
polarity over small scales, reconnects and accelerates the flow gradually
(Drenkhahn 2002; Drenkhahn \& Spruit 2002). Assuming that a large fraction of the
dissipated magnetic energy heats the electrons in the flow, both analytical
estimates and  numerical calculations show that at a radius that lies a factor of
ten below the Thomson photosphere $r_{\rm{ph}}$ radiation and matter
are no-longer is thermal equilibrium. As a result of the energy dissipation, the 
flow develops a ``hot'' photosphere where electron temperature increases as a 
function of distance with comoving temperatures $\sim$ a few tens keV at the location 
of the photosphere. This result is rather independent of the model parameters such as the 
luminosity of the flow or the baryon loading $\sigma_0$.
 
Inverse Compton scattering plays an important role at the photosphere, leading to Comptonization
spectra that have characteristic non-thermal appearance. Fits of the numerically calculated
spectra with the Band spectrum (Band et al. 1993)
give parameters of the low/high frequency slope and the $E_{\rm{peak}}$ of the spectrum in
$E\cdot f(E)$ representation in agreement with observations.
Furthermore, the observed clustering of the  $E_{\rm{peak}}$ in the sub-MeV range is a natural outcome
of the model. The strength of the photospheric component is $\sim$3-20\% 
that of the luminosity of the flow and has most ($\simmore 60$\%) of its energy
in the BATSE energy range. 
I, therefore, conclude that the photospheric component expected from magnetic models 
can to a large extent be responsible for the prompt GRB emission.


\begin{acknowledgements}
I thank Henk Spruit for many valuable suggestions and discussions and for 
carefully reading the manuscript. 
I acknowledge support from the EU FP5 Research Training Network ``Gamma Ray Bursts:
An Enigma and a Tool.'' 
\end{acknowledgements}


\begin{thebibliography}{}

\bibitem{} Abramowicz, M. A., Novikov, I. D., \& Paczy\'nski, B. 1991, ApJ, 369, 175
\bibitem{} Band, D., Matteson, J., Ford, L., et al. 1993, ApJ, 413, 281
\bibitem{} Barraud, C., Olive, J. -F., Lestrade, J. P. et al. 2003, A\&A, 400, 1021
\bibitem{} Coroniti, F. V. 1990, ApJ, 349, 538
\bibitem{} Crider, A., Liang, E. P., Smith, I. A., et al. 1997, ApJ, 479, L39
\bibitem{} Daigne, F., \& Mochkovitch, R. 2002, MNRAS, 336, 1271 
\bibitem{} Drenkhahn, G. 2002, A\&A, 387, 714
\bibitem{} Drenkhahn, G., \& Spruit, H. C. 2002, A\&A, 391, 1141
\bibitem{} Frontera, F., Amati, L., Costa, E., et al. 2000, ApJS, 127, 59
\bibitem{} Ghirlanda, G., Celotti, A., \& Ghisellini, G. 2003, A\&A, 406, 879
\bibitem{} Ghisellini, G., \& Celotti, A. 1999, A\&AS, 138, 527
\bibitem{} Giannios, D., \& Spruit, H. C. 2005, A\&A, 430, 1
\bibitem{} Giannios, D., \& Spruit, H. C. 2006, A\&A, 450, 887
\bibitem{} Goodman, J. 1986, ApJ, 308, L47
\bibitem{} Granot, J., Ramirez-Ruiz, E., \& Perna, R. 2005, ApJ, 630, 1003
\bibitem{} Liang, E. P. 1997, ApJ, 491, L15
\bibitem{} Lightman, A. P. 1982, ApJ, 253, 842
\bibitem{} Lyubarsky, Y. E. 2005, MNRAS, 358, 113
\bibitem{} Lyutikov, M, \& Uzdensky, D. 2003, ApJ, 589, 893
\bibitem{} Lyutikov, M., \& Blandford R. D. 2003, astro-ph/0312347
\bibitem{} M\'esz\'aros, P., \& Rees, M. J. 1997, ApJ, 482, L29
\bibitem{} M\'esz\'aros, P., \& Rees, M. J. 2000, ApJ, 530, 292
\bibitem{} Paczy\'nski, B. 1986, ApJ, 308, L43
\bibitem{} Pe'er, A., \& Waxman, E. 2004, ApJ, 613, 448 
\bibitem{} Pe'er, A., M\'esz\'aros, P., \& Rees, M. J. 2005, ApJ, 635, 476
\bibitem{} Pe'er, A., M\'esz\'aros, P., \& Rees, M. J. 2006, ApJ, 642, 995
\bibitem{} Petrosian, V. 1981, ApJ, 251, 727
\bibitem{} Petrosian, V., \& McTiernan, J. M. 1983, Phys. Fluids, 26, 3023  
\bibitem{} Piran, T. 1999, Phys. Rep., 314, 575
\bibitem{} Pozdniakov, L. A., Sobol, I. M., Sunyaev, R. A. 1983, Astrophysics \& Space 
Physics Rev. 2, 189
\bibitem{} Preece, R. D., Briggs, M. S., Mallozzi, R. S, et al. 1998, ApJ, 506, L23
\bibitem{} Ramirez-Ruiz, E. 2005, MNRAS, 363, L61
\bibitem{} Rees, M. J., \& M\'esz\'aros, P. 1994, ApJ, 430, L93
\bibitem{} Rees, M. J., \& M\'esz\'aros, P. 2005, ApJ, 628, 847
\bibitem{} Rybicki, G. B. , \& Lightman, A. P. 1979, {\emph Radiative Processes in 
Astrophysics}, Wiley, New York
\bibitem{} Ryde, F. 2004, ApJ, 614, 827
\bibitem{} Ryde, F. 2005, ApJ, 625, L95
\bibitem{} Sari, R., \& Piran, T. 1997, MNRAS, 287, 110
\bibitem{} Spitzer, L. 1956, {\emph Physics of Fully Ionized Gases}, Wiley, New York
\bibitem{} Spruit, H. C., Daigne, F., \& Drenkhahn, G. 2001, A\&A, 369, 694
\bibitem{} Stepney, S. 1983, MNRAS, 202, 467
\bibitem{} Svensson, R. 1982, ApJ, 258, 335
\bibitem{} Thompson, C. 1994, MNRAS, 270, 480
\bibitem{} Vlahakis, N., \& K\"onigl, A. 2003, ApJ, 596, 1104
\bibitem{} Wardzi\'nski, G., \& Zdziarski, A. A. 2000, MNRAS, 314, 183

\end{thebibliography}
\end{document}